# Spin-Phonon Interaction in Yttrium Iron Garnet

Kevin S. Olsson[1†], Jeongheon Choe[1,2], Martin Rodriguez-Vega[1,2,3*], Guru Khalsa[4], Nicole A. Benedek[4], Bin Fang[1,2], Jianshi Zhou[2,5,6], Gregory A. Fiete[1,2,3,7], and Xiaoqin Li[1,2,6*]

1. Department of Physics, Center of Complex Quantum Systems, The University of Texas at Austin, Austin, Texas 78712, USA
2. Center for Dynamics and Control of Materials, The University of Texas at Austin, Austin, Texas 78712, USA
3. Department of Physics, Northeastern University, Boston, MA 02115, USA
4. Department of Materials Science and Engineering, Cornell University, Ithaca, New York 14853, USA
5. Department of Mechanical Engineering, The University of Texas at Austin, Austin, Texas 78712, USA
6. Texas Materials Institute, The University of Texas at Austin, Austin, Texas 78712, USA
7. Department of Physics, Massachusetts Institute of Technology, Cambridge, MA 02139, USA

* Corresponding Authors: elaineli@physics.utexas.edu, rodriguezvega@utexas.edu

† Current Address: Department of Computer and Electrical Engineering, University of Maryland, College Park, MD 20740, USA

**Abstract**: Spin-phonon interaction is an important channel for spin and energy relaxation in magnetic insulators. Understanding this interaction is critical for developing magnetic insulator-based spintronic devices. Quantifying this interaction in yttrium iron garnet (YIG), one of the most extensively investigated magnetic insulators, remains challenging because of the large number of atoms in a unit cell. Here, we report temperature-dependent and polarization-resolved Raman measurements in a YIG bulk crystal. We first classify the phonon modes based on their symmetry. We then develop a modified mean-field theory and define a symmetry-adapted parameter to quantify spin-phonon interaction in a phonon-mode specific way for the first time in YIG. Based on this improved mean-field theory, we discover a positive correlation between the spin-phonon interaction strength and the phonon frequency.

## Introduction

Magnetic insulators are of considerable interest in spintronics due to their minimal spin damping [1–3]. This low damping originates in part from the absence of low-energy electronic excitations, leaving the spins to interact primarily with other spins (magnons) and the lattice (phonons). Beyond their role in spin excitation damping, interactions between the magnons and phonons play a crucial role in developing devices based on thermally driven spin transport [4–6], spin pumping through hybrid spin-lattice excitations[7], and magnon cavity quantum electrodynamics [8, 9]. Of various magnetic insulators explored for spintronic devices, yttrium iron garnet (YIG): $Y_3Fe_5O_{12}$ is the most widely investigated due to its remarkably low spin damping and its high transition temperature of 560 K [10, 11]. However, due to its massive unit cell (160 atoms as an inset of Fig.1a), extracting the spin-phonon interaction (SPI) of YIG from *ab initio* studies is remarkably difficult.

The SPI in YIG has been investigated through different types of experiments. Brillouin light scattering and spin Seebeck transport measurements of YIG have examined the interactions of magnons and phonons through quasiparticle hybridization [12–14]. Other studies have touched upon the SPI by measuring the magnon-phonon energy relaxation length and time [4, 15]. However, no study provides a direct and quantitative measurement of the strength of the SPI in YIG in a phonon-mode specific way.

Without knowing the SPI strength, it is difficult to develop accurate models of spin relaxation in YIG or compare YIG to other magnetic insulators for device development.

Here we report Raman spectroscopy studies of optical phonons in a YIG bulk crystal. By analyzing their symmetry properties and temperature-dependent phonon frequency shift, we investigate if SPI changes systematically for each phonon mode. We determine that the complex unit cell precludes a direct correlation between symmetry or frequency of a phonon mode with the conventional $\lambda$-model of the SPI strength [16–18]. By developing a mean-field model and defining a new parameter to describe SPI strength, we observe a correlation between this mean-field SPI parameter and phonon frequency. These results provide crucial information and advance the understanding of how magnons and phonons interact in YIG.

**Experiment**

YIG ($Y_3Fe_5O_{12}$) is an insulating ferrimagnet (FiM) with Curie temperature $T_C$ = 570 K and easy axis in the [111] direction [19, 20]. YIG crystals exhibit symmetries described by cubic space group $Ia\bar{3}d$ (No. 230) and point group $O_h$ at the Γ point [21–23]. Inversion symmetry present in $O_h$ implies that the phonon modes show mutually-exclusive infrared and Raman activity. The possible Raman irreducible representations in $O_h$ are either $T_{2g}$, $E_g$, or $A_{1g}$. The crystal structure is composed of Y atoms occupying the 24c Wyckoff sites, Fe ions in the 16a and 24d positions, and O atoms in the 96h sites. The conventional unit cell has eight formula units, with 24 Y ions, 40 Fe ions, and 96 O ions for a total of 160 atoms.

Raman measurements were taken with a 532 nm laser incident on a YIG single crystal with [111] oriented along the surface normal. The scattered light was collected in a backscattering geometry and directed to a diffraction grating-based spectrometer. The observed optical phonon modes in the Raman spectra agree with previous measurements of YIG [24, 25]. Low-temperature measurements from 8.8 K to 313.65 K were performed in a closed-loop cryostat, and high-temperature measurements from 313.65 K to 631.95 K were performed with a ceramic heater. Between each temperature, the sample was allowed to equilibrate for 15 minutes or longer. A saturating magnetic field was applied in the sample plane for all measurements. Due to constraints of the experiment systems, low-temperature measurements used a 300 mT saturating field, and the high-temperature measurements used a 50 mT saturating field. Since both fields were above the saturating field, this difference did not noticeably affect the magnetic ordering of YIG or the Raman spectra.

Raman spectra were collected with a fixed *s*-polarization incident on the sample. Fig.1a shows the *p*- and *s*-polarizations components of the scattered light for the sample at low temperature (8.8 K). Phonon modes of different symmetries scatter light with different polarizations. Fig. 1(b) shows the intensity of the Raman signal from the scattered light as it passed through a linear polarizer, with the polarization axis rotated in steps of 20° from 0° to 180°. Based on the results, the phonons are categorized with their respective irreducible representations: $T_{2g}$, $E_g$, or $A_{1g}$.

The temperature dependence of the phonon frequencies was determined by fitting with a Lorentzian function and extracting the central frequencies. We plot the measured Raman spectra for one $T_{2g}$ mode at three different representative temperatures 8.8 K, 313.65 K, and 632 K in Fig. 2 (a), (b), and (c). At low temperatures (e.g. 8.8 K), the low thermal population of the phonons reduces the Raman intensity. In contrast, the phonon modes exhibit a broader linewidth at high temperatures due to increased phonon-phonon and phonon-magnon scattering, which lowers the peak intensity. Consequently, the temperature-dependent frequency was only measurable for a subset of the observed phonons. The temperature dependence of the peak frequencies for the two modes is shown in Fig. 2d and 2e. The temperature dependence of peak frequencies of all the measurable phonon modes can be found in Supplementary Information.

**Results**

In the absence of spin order above the transition temperature (i.e. 559 K for YIG), the temperature dependence of the optical phonon frequency $\omega_p$ is determined by anharmonic effects, i.e. phonon-phonon scattering. Well below the melting points, 3-phonon scattering dictates the temperature dependence of $\omega_p$ as follows

$$\omega_p(T) = \omega_p(0) - A\left(1 + \frac{2}{\mathrm{Exp}[x] - 1}\right) \quad (1)$$

where $\omega_p(0)$ is the zero temperature phonon frequency, $A$ is a coefficient related to the 3-phonon scattering strength and $x = \hbar\omega_p(0)/2k_BT$ with Planck's constant $\hbar$, Boltzmann's constant $k_B$, and temperature $T$ [16–18]. We fit the peak frequency above 559 K using Eq. (1) to determine $\omega_p(T)$ for each phonon mode. Examples of these fits are shown in Fig. 2d and 2e.

In the magnetically ordered state, the influence of spin order on the phonon frequency can be treated as a small deviation, $\Delta\omega_{sp}$, such that the optical phonon frequency is given as

$$\omega_p' = \omega_p(T) + \Delta\omega_{sp} \quad (2)$$

where $\omega_p'$ is the measured phonon frequency. Then, $\Delta\omega_{sp}$ can be found by taking the difference of the measured frequency and anharmonic temperature-dependent phonon frequency, i.e. $\omega_p' - \omega_p(T)$, as shown in Fig. 3 for selected phonon modes.

Many previous studies of the SPI express the frequency deviation as $\Delta\omega_{sp} = \lambda\langle \boldsymbol{S}_i \cdot \boldsymbol{S}_j \rangle$, where $\lambda$ is a single term capturing the SPI strength and $\langle \boldsymbol{S}_i \cdot \boldsymbol{S}_j \rangle$ represents nearest-neighbor spin correlation function [26–30]. The spin correlation function can be approximated $\langle \boldsymbol{S}_i \cdot \boldsymbol{S}_j \rangle \approx S_z^2 B_J(T)$, where $B_J$ is the Brillouin function, which has a maximum value of 1 at $T/T_c = 0$ [27]. Thus to find $\lambda$ without the spin-related, temperature-dependent contribution to the frequency, $\Delta\omega_{sp}$ should be evaluated at $T = 0$. Table 1 reports frequency deviation measured at 8.8 K, $\Delta\omega_{sp}^0 = \omega_p' - \omega_p(8.8\ \mathrm{K})$, the lowest temperature reached in our experiments. The high $T_c$ of YIG and slow decrease of $B_J(T)$ results in $B_J(8.8\ \mathrm{K}) \approx 1$. Then, $\langle \boldsymbol{S}_i \cdot \boldsymbol{S}_j \rangle \approx S_z^2$ and using $S_z = \frac{5}{2}$ for the magnetic iron atoms in YIG, $\lambda$ is found from $\Delta\omega_{sp}^0$, also reported in Table 1. Examining the results shown in Table 1, there is no clear trend for $\Delta\omega_{sp}^0$ and $\lambda$ with either the frequency or symmetry of the mode. These results highlight the deficiency of the $\lambda$ model that has been applied successfully for other materials with a simple unit cell such as $FeF_2$ and $ZnCr_2O$[26–30].

**Discussion**

The simple $\lambda$ model, which treats all phonon modes equally, is insufficient for describing the SPI in YIG. This is not surprising as the large unit cell leads to complicated phonon dispersions. However, a more detailed first-principles approach like density functional theory (DFT) for determining the SPI is exceedingly difficult, again due to the large unit cell of YIG, as well as the especially high precision required in the computations to accurately describe the lattice vibrations and their coupling to magnetic order. Thus, to describe spin-phonon interaction in YIG, we develop a modified mean-field model that captures the mode dependence of the SPI.

We begin with the Ginzburg-Landau (GL) potential describing the magnetic order,

$$F = \frac{A}{2}m^2 + \frac{B}{2}m^4 \quad (3)$$

where $m \equiv M/M_0$ is the ferrimagnetic order parameter defined as the magnetization ($M$) divided by its zero temperature value ($M_0$). The GL parameters $A$ and $B$ have units of energy and $A = -a(T_c - T)$, where $T_c$ is the magnetic transition temperature. The temperature dependence of the order parameter agrees well with the temperature dependence of the magnetic moment of YIG reported in the literature (Supplementary Information).

This GL potential only includes magnetic order, and thus needs to be expanded to include phonon contribution to the GL potential. By including only the harmonic terms, the GL potential takes the form

$$F = \frac{A}{2}m^2 + \frac{B}{2}m^4 + \frac{1}{2}\mu\omega u^2 + \frac{1}{2}\delta_{sp}m^2u^2 \quad (4)$$

where $\mu$ is the phonon mode reduced mass, $u$ is the atomic displacement, and $\delta_{sp}$ is the SPI strength [31, 32]. Note that for phonons with irreducible representation $A_g$ and $T_{2g}$, the symmetry allows a cubic term proportional to $m^2u$, which is weak in YIG (see Supplementary Information) [33].

Equilibrium values $m_*$ and $u_*$ are found from the conditions

$$\frac{\partial F}{\partial m} = 0, \frac{\partial F}{\partial u} = 0 \quad (5)$$

and the spin-dependent phonon frequency ($\Omega$) is determined by

$$\mu\Omega^2 = \left.\frac{\partial^2 F}{\partial u^2}\right|_{\substack{m=m_* \\ u=u_*}} = \mu\omega^2 + \delta_{sp}m_*^2 \,. \quad (6)$$

Now, using the equilibrium value of $m_* = \sqrt{a(T_c - T)/B}$, $\Omega$ is approximately given by

$$\Omega(T) \approx \omega + \frac{\delta_{sp}}{2\mu\omega}\left(1 - \frac{T}{T_c}\right) \quad (7)$$

to first order in $\delta_{sp}$. (Note: $\delta_{sp}$ is defined for angular frequencies.) Compared to the $\lambda$ model, we see that the frequency deviation is determined by the frequency and reduced mass of the phonon mode, as well as the SPI strength. We use this improved mean-field theory to extract the SPI strength. Figure 3 shows $\Delta\omega_{sp}$ across the temperature range, with fits using Eq. (7) to extract the $\delta_{sp}$, shown in Figure 4.

To further understand the SPI found from the modified mean-field model, we examine the atomic displacements of each phonon mode. Using group theory projection operators, we can derive a basis of eigenmodes that brings the dynamical matrix to a block-diagonal form[34]. The 739 $cm^{-1}$ mode only involves the O atoms' displacements due to its $A_g$ symmetry (see Supplementary Information). Because it only involves O atoms, this mode has the smallest reduced mass μ compared with $T_{2g}$ and $E_g$ modes. We find that this phonon mode has the largest $\delta_{sp}$. This finding is consistent with the interpretation that the vibrations of the light O atoms are most affected by the magnetic ordering of the heavy Fe atoms. The symmetries of other phonon modes, $T_{2g}$ and $E_g$, allow motions of all three ion types (Y, Fe, O) in principle. First-principles calculations of the Raman phonon frequencies and symmetries allow us to assign $\mu$ to each Raman phonon. We find that, as expected, lower frequency phonons have larger $\mu$. (See Supplementary Information.) Using these values of $\mu$ to calculate the SPI, we find that higher frequency phonons have larger SPI as shown in Figure 4. This trend suggests that the atoms with stronger bonds (consequently higher phonon frequency) are more affected by magnetic ordering.

**Conclusion**

In summary, we investigate SPI associated with optical phonon modes of a YIG bulk crystal. By taking polarization-resolved Raman spectra, we analyze their symmetry. Temperature-dependent Raman spectra taken over a broad temperature range of 8.8-635 K allow us to evaluate SPI quantitatively and specific to a particular phonon mode. By developing an improved mean-field model and applying a refined analysis, we discover that the SPI increases with phonon frequency. The $A_g$ mode involving vibrations of only O atoms has the strongest SPI. These results provide both direct and mode-specific interaction strengths, thus, providing valuable information for advancing theories of magnetic insulators and for exploring spintronic devices such as those based on spin-caloritronic effects.

**Acknowledgements**

This research was primarily supported (J.C., M.R.-V., B.F., J.Z., G.A.F., X.L.) by the National Science Foundation through the Center for Dynamics and Control of Materials: an NSF MRSEC under Cooperative Agreement No. DMR1720595. G.K. and N.A.B. were supported by the Cornell Center for Materials Research: an NSF MRSEC under Cooperative Agreement No. DMR-1719875. G.A.F also acknowledges support from NSF Grant No. DMR-1949701.

K.S.O. and J.C. performed measurements. B.F. built the experimental system. K.S.O. and J.C. analyzed experimental results. M.R.-V. and G.A.F. developed mean-field model and performed group theory analysis with input from G.K. G.K. and N.A.B. performed first-principles evaluation. J.Z. provided the YIG sample. K.S.O., J.C., M.R.-V., and X.L. wrote the original manuscript. X.L. and G.A.F. supervised the project. All authors assisted in the discussion of results and revision of the manuscript.

**References**

[1] V. V Kruglyak, S. O. Demokritov and D. Grundler, “Magnonics,” *J. Phys. D. Appl. Phys.*, vol. **43**, no. 26, p. 264001, Jul. 2010.

[2] A. A. Serga, A. V. Chumak and B. Hillebrands, “YIG magnonics,” *J. Phys. D. Appl. Phys.*, vol. **43**, no. 26, p. 264002, 2010.

[3] A. V. Chumak and H. Schultheiss, “Magnonics: Spin waves connecting charges, spins and photons,” *Journal of Physics D: Applied Physics*, vol. **50**, no. 30. p. 300201, 2017.

[4] A. Prakash, B. Flebus, J. Brangham, F. Yang, Y. Tserkovnyak and J. P. Heremans, “Evidence for the role of the magnon energy relaxation length in the spin Seebeck effect,” *Phys. Rev. B*, vol. **97**, no. 2, p. 020408, Jan. 2018.

[5] L. J. Cornelissen, K. J. H. Peters, G. E. W. Bauer, R. A. Duine and B. J. Van Wees, “Magnon spin transport driven by the magnon chemical potential in a magnetic insulator,” *Phys. Rev. B*, vol. **94**, no. 1, p. 014412, 2016.

[6] K. S. Olsson, K. An, G. A. Fiete, J. Zhou, L. Shi and X. Li, “Pure Spin Current and Magnon Chemical Potential in a Nonequilibrium Magnetic Insulator,” *Phys. Rev. X*, vol. **10**, no. 2, p. 021029, Jun. 2020.

[7] H. Hayashi and K. Ando, “Spin Pumping Driven by Magnon Polarons,” *Phys. Rev. Lett.*, vol. **121**, no. 23, p. 237202, 2018.

[8] X. Zhang, C. L. Zou, L. Jiang and H. X. Tang, “Cavity magnomechanics,” *Sci. Adv.*, vol. **2**, no. 3, p. 1501286, 2016.

[9] J. Li, S. Y. Zhu and G. S. Agarwal, “Magnon-Photon-Phonon Entanglement in Cavity Magnomechanics,” *Phys. Rev. Lett.*, vol. **121**, no. 20, p. 203601, 2018.

[10] V. Cherepanov, I. Kolokolov and V. L’vov, “The saga of YIG: Spectra, thermodynamics, interaction and relaxation of magnons in a complex magnet,” *Phys. Rep.*, vol. **229**, no. 3, pp. 81–144, Jul. 1993.

[11] A. Prabhakar and D. D. Stancil, *Spin waves: Theory and applications*. New York, NY: Springer,

2009.

[12] T. Kikkawa, K. Shen, B. Flebus, R. A. Duine, K. Uchida, Z. Qiu, G. E. W. Bauer and E. Saitoh, “Magnon Polarons in the Spin Seebeck Effect,” *Phys. Rev. Lett.*, vol. **117**, no. 20, p. 207203, Nov. 2016.

[13] D. A. Bozhko, A. A. Serga, P. Clausen, V. I. Vasyuchka, F. Heussner, G. A. Melkov, A. Pomyalov, V. S. L’Vov and B. Hillebrands, “Supercurrent in a room-temperature Bose-Einstein magnon condensate,” *Nat. Phys.*, vol. **12**, no. 11, pp. 1057–1062, 2016.

[14] H. Man, Z. Shi, G. Xu, Y. Xu, X. Chen, S. Sullivan, J. Zhou, K. Xia, J. Shi and P. Dai, “Direct observation of magnon-phonon coupling in yttrium iron garnet,” *Phys. Rev. B*, vol. **96**, no. 10, 2017.

[15] S. F. Maehrlein, I. Radu, P. Maldonado, A. Paarmann, M. Gensch, A. M. Kalashnikova, R. V. Pisarev, M. Wolf, P. M. Oppeneer, J. Barker and T. Kampfrath, “Dissecting spin-phonon equilibration in ferrimagnetic insulators by ultrafast lattice excitation,” *Sci. Adv.*, vol. **4**, no. 7, p. aar5164, 2018.

[16] R. A. Cowley, “Anharmonic crystals,” *Reports Prog. Phys.*, vol. **31**, no. 1, pp. 123–166, Jan. 1968.

[17] M. Balkanski, R. F. Wallis and E. Haro, “Anharmonic effects in light scattering due to optical phonons in silicon,” *Phys. Rev. B*, vol. **28**, no. 4, pp. 1928–1934, 1983.

[18] J. Menéndez and M. Cardona, “Temperature dependence of the first-order Raman scattering by phonons in Si, Ge, and -Sn: Anharmonic effects,” *Phys. Rev. B*, vol. **29**, no. 4, pp. 2051–2059, 1984.

[19] S. Geller and M. A. Gilleo, “The crystal structure and ferrimagnetism of yttrium-iron garnet, Y3Fe2(FeO4)3,” *J. Phys. Chem. Solids*, vol. **3**, no. 1–2, pp. 30–36, Jan. 1957.

[20] S. Geller and M. A. Gilleo, “The effect of dispersion corrections on the refinement of the yttrium-iron garnet structure,” *J. Phys. Chem. Solids*, vol. **9**, no. 3–4, pp. 235–237, Mar. 1959.

[21] G. P. Rodrigue, H. Meyer and R. V. Jones, “Resonance measurements in magnetic garnets,” *J. Appl. Phys.*, vol. **31**, no. 5, pp. S376–S382, May 1960.

[22] A. B. Harris, “Spin-wave spectra of yttrium and gadolinium iron garnet,” *Phys. Rev.*, vol. **132**, no. 6, pp. 2398–2409, Dec. 1963.

[23] M. Wu, “Nonlinear spin waves in magnetic film feedback rings,” in *Solid State Physics - Advances in Research and Applications*, vol. **62**, Academic Press, 2010, pp. 163–224.

[24] S. Khanra, A. Bhaumik, Y. D. Kolekar, P. Kahol and K. Ghosh, "Structural and magnetic studies of Y3Fe5-5xMo 5xO12," *J. Magn. Magn. Mater.*, vol. **369**, pp. 14–22, 2014.

[25] J. M. Costantini, S. Miro, F. Beuneu and M. Toulemonde, "Swift heavy ion-beam induced amorphization and recrystallization of yttrium iron garnet," *J. Phys. Condens. Matter*, vol. **27**, no. 49, p. 496001, 2015.

[26] C. J. Fennie and K. M. Rabe, "Magnetically induced phonon anisotropy in ZnCr2O4 from first principles," *Phys. Rev. Lett.*, vol. **96**, no. 20, p. 205505, May 2006.

[27] D. J. Lockwood and M. G. Cottam, "The spin-phonon interaction in FeF2 and MnF2 studied by Raman spectroscopy," *J. Appl. Phys.*, vol. **64**, no. 10, pp. 5876–5878, 1988.

[28] A. B. Sushkov, O. Tchernyshyov, W. Ratcliff, S. W. Cheong and H. D. Drew, "Probing spin correlations with phonons in the strongly frustrated magnet ZnCr2O4," *Phys. Rev. Lett.*, vol. **94**, no. 13, p. 137202, Apr. 2005.

[29] D. J. Lockwood, "Spin-phonon interaction and mode softening in NiF2," *Low Temp. Phys.*, vol. **28**, no. 7, pp. 505–509, Jul. 2002.

[30] E. Aytan, B. Debnath, F. Kargar, Y. Barlas, M. M. Lacerda, J. X. Li, R. K. Lake, J. Shi and A. A. Balandin, "Spin-phonon coupling in antiferromagnetic nickel oxide," *Appl. Phys. Lett.*, vol. **111**, no. 25, p. 252402, Dec. 2017.

[31] D. Dey, T. Maitra, U. V. Waghmare and A. Taraphder, "Phonon dispersion, Raman spectra, and evidence for spin-phonon coupling in MnV2 O4 from first principles," *Phys. Rev. B*, vol. **101**, no. 20, p. 205132, May 2020.

[32] A. Paul, P. Sharma and U. V. Waghmare, "Spin-orbit interaction, spin-phonon coupling, and anisotropy in the giant magnetoelastic effect in YMnO3," *Phys. Rev. B*, vol. **92**, no. 5, p. 054106, Aug. 2015.

[33] R. Z. Levitin, A. S. Markosyan and V. N. Orlov, "X-ray study of crystalline structure magnetoelastic distortions in Tb3−xYxFe5O12 terbium-yttrium ferrite-garne," *Sov. Phys. Solid State*, vol. **25**, no. 6, pp. 1074–1075, 1983.

[34] M. S. Dresselhaus, G. Dresselhaus and A. Jorio, *Group theory*. Springer Berlin Heidelberg, 2008.

| Phonon frequency (cm$^{-1}$) | Symmetry | Spin-phonon frequency deviation $\Delta\omega_{sp}^{0}$ (cm$^{-1}$) | Coupling strength λ (cm$^{-1}$) |
|---|---|---|---|
| 174 | $T_{2g}$ | 8.4 ± 0.9 | 1.3 ±0.1 |
| 194 | $T_{2g}$ | 5.1 ± 0.8 | 0.8 ±0.1 |
| 239 | $T_{2g}$ | 11.4 ± 1.0 | 1.8 ±0.2 |
| 276 | $E_g$ | 3.5 ± 0.6 | 0.6 ±0.1 |
| 346 | $E_g$ | 12± 2 | 1.9 ±0.4 |
| 378 | $T_{2g}$ | -2.3 ± 4.2 | -0.4 ±0.7 |
| 447 | $T_{2g}$ | 8 ± 4 | 1.3 ±0.6 |
| 508 | $E_g$ | 8 ± 2 | 1.3 ±0.3 |
| 591 | $T_{2g}$ | 10 ± 3 | 1.6 ±0.5 |
| 739 | $A_{1g}$ | 11 ± 5 | 1.8 ±0.8 |

TABLE 1. Symmetry, spin-phonon frequency deviation, and λ coupling strength of the measured phonon modes in YIG.

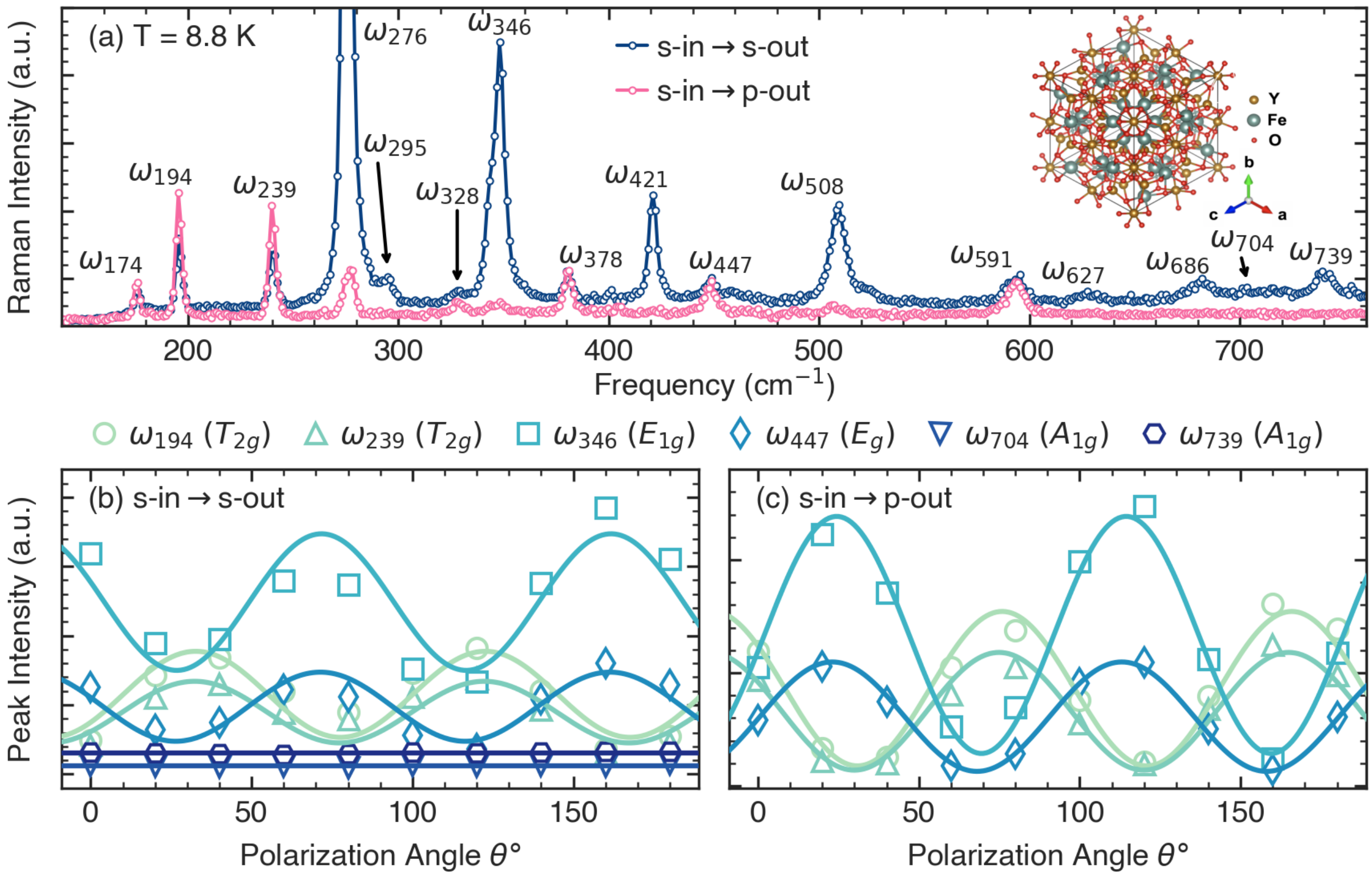

FIG. 1. (a) Raman spectra taken with s-in/s-out (colinear) and s-in/p-out (crossed) polarization configurations at 8.8 K. Solid lines connect data points for clarity. Inset shows the YIG crystal structure along the [111] direction. (b,c) Angle-dependent intensities of the representative $A_{1g}$, $E_g$, and $T_{2g}$ modes. The spectra were obtained by by keeping incident polarization fixed. Panel b and c refer to colinear and crossed polarization configurations, respectively. The fit curves follow theoretical predictions from crystal lattice symmetry.

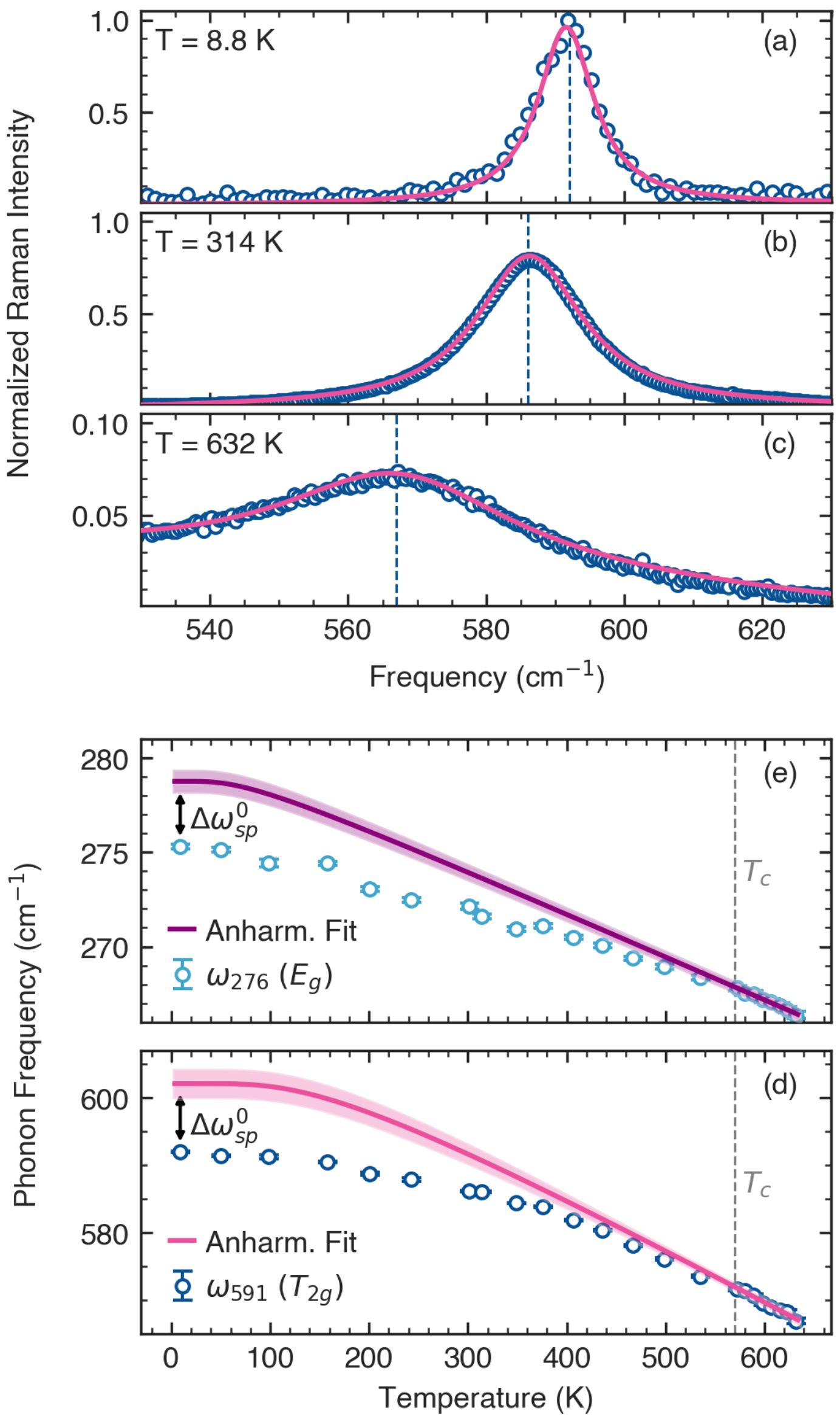

FIG. 2. (a,b,c) Example spectra for different temperatures, normalized to the peak intensity at 8.8 K. Solid lines are Lorentzian fits with a linear offset to account for the background. Vertical dashed lines indicate the peak positions. (d,e) Temperature dependence of $\omega_{276}$ and $\omega_{591}$ phonon frequencies, which have symmetries $T_{2g}$ and $E_g$, respectively. The solid curves correspond to the anharmonic phonon-phonon scattering fit, which is based on fitting to data only above the temperature $T_c$. The deviation from the anharmonic curve (black arrow) reflects the corresponding spin-phonon coupling strength, λ, given in Eq.(2).

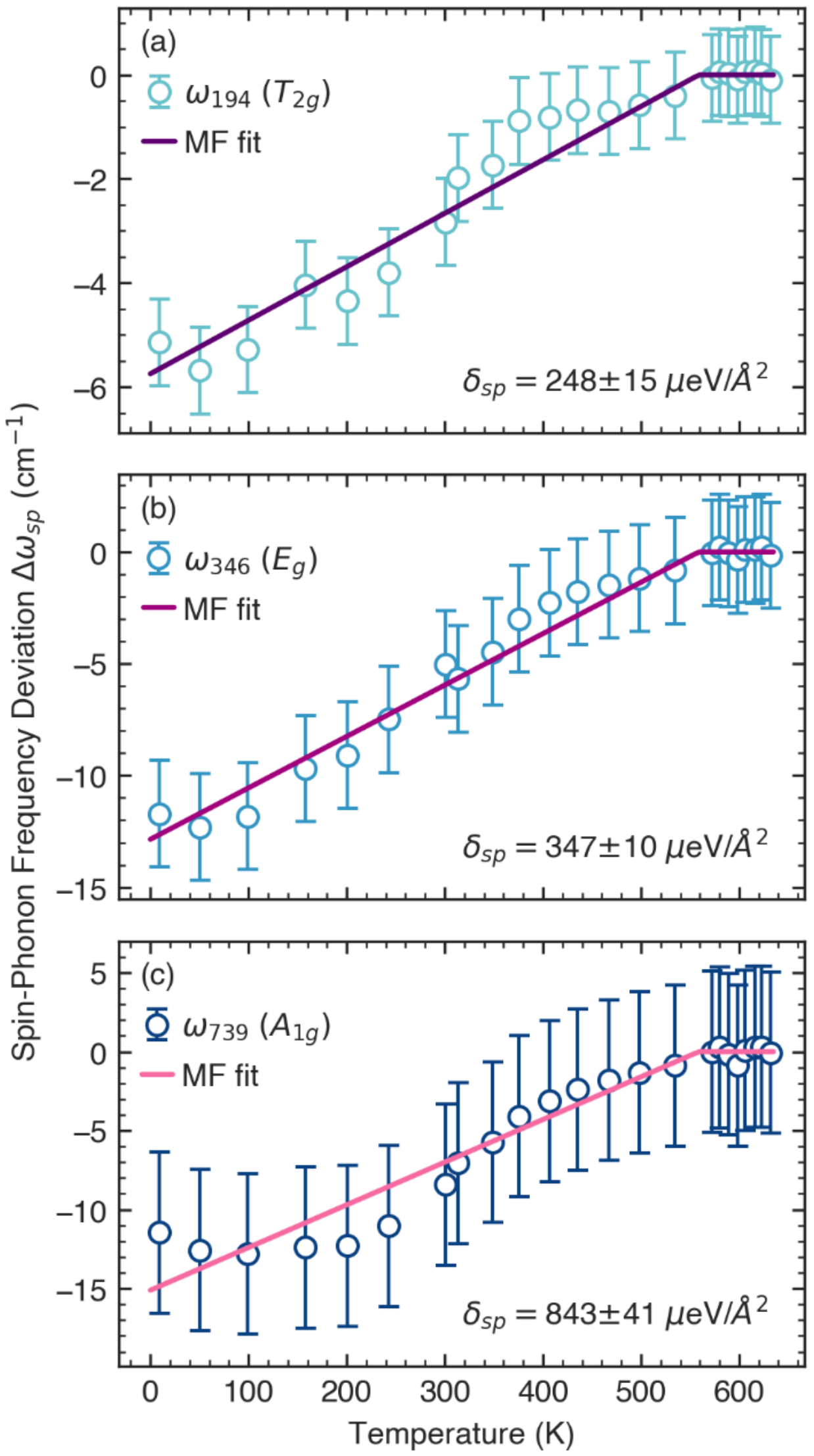

FIG. 3. The measured phonon frequencies is subtracted from the temperature-dependent frequency found with the anharmonic fit [see Fig. 2 (b) and (c)] to determine $\Delta\omega_{sp}$. Solid lines show fits to the mean-field model which yield the spin-phonon interaction strength $\delta_{sp}$, given in Eq.(7).

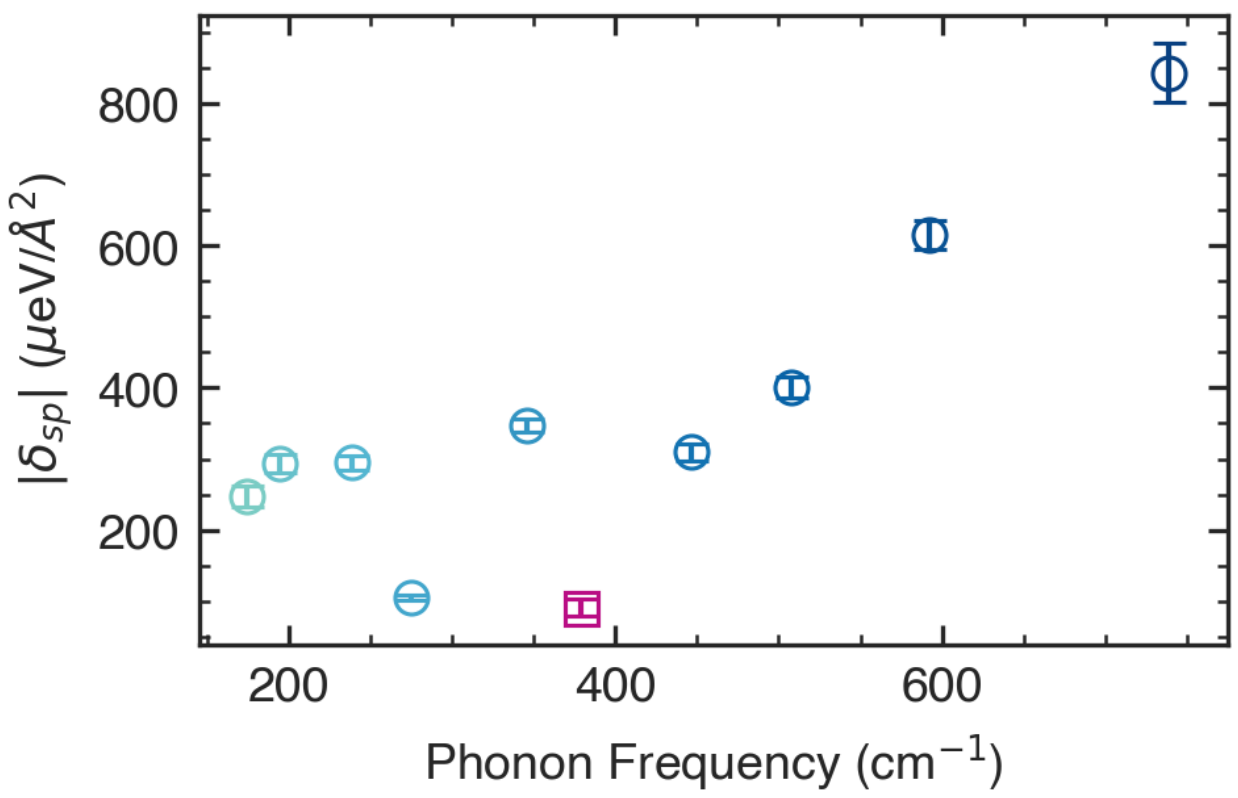

FIG. 4. Absolute value of the spin-phonon interaction strength evaluated with the mean-field model for the phonon modes in YIG. The measured $\delta_{sp}$ for the $\omega_{378}$ mode is negative (purple square), while the rest of the measured $\delta_{sp}$ are positive.

# Supplementary Information: Spin-Phonon Interaction in Yttrium Iron Garnet

Kevin S. Olsson[1†], Jeongheon Choe[1,2], Martin Rodriguez-Vega[1,2,3]*, Guru Khalsa[4], Nicole A. Benedek[4], Bin Fang[1,2], Jianshi Zhou[2,5,6], Gregory A. Fiete[1,2,3,7], and Xiaoqin Li[1,2,6]*

**Table of Contents**

## Section 1: Group theory aspects and Raman activity

YIG ($Y_3Fe_5O_{12}$) is an insulating ferrimagnet (FiM) with Curie temperature $T_C$ = 570 K, cubic space group $Ia\bar{3}d$ (No. 230), and point group $O_h$ at the Γ point [1, 2]. The crystal structure is composed of Y atoms occupying the 24c Wyckoff sites, Fe ions in the 16a and 24d positions, and O atoms in the 96h sites. The conventional unit cell has eight formula units, with 24 Y ions, 40 Fe ions, and 96 O ions. Typically, the easy axis is in the [111] direction [3–5]. We start our group theory analysis by determining the Raman activity of the crystal. First, we note that in the $O_h$ point group, the representation of the vector is $\Gamma_{\text{vec.}} = T_{1u}$. Then, we calculate the equivalence representation employing the Bilbao crystallographic server [6]. The equivalence representation is the number of atoms that remain invariant under the point group symmetry operations for each irreducible representation and is given by

$$\Gamma_{\text{eq.}} = 4A_{1g} + 2A_{1u} + 2A_{2g} + 2A_{2u} + 4E_u + 4E_g + 5T_{2u} + 5T_{2g} + 3T_{1u} + 5T_{1g}. \tag{S1}$$

Then, the representation of the lattice vibrations is

$$\begin{aligned}\Gamma_{\text{lat. vib.}} &= \Gamma_{\text{eq.}} \otimes \Gamma_{\text{vec.}} \\ &= 3A_{1g} + 5A_{2g} + 8E_g + 14T_{1g} + 14T_{2g} + 5A_{1u} + 5A_{2u} + 10E_u + 18T_{1u} + 16T_{2u},\end{aligned} \tag{S2}$$

which contains the acoustic modes with symmetry $T_{1u}$. Therefore, we expect 25 (first-order) Raman active modes:

$$\Gamma_{\text{Raman}} = 3A_g \oplus 8E_g \oplus 14T_{2g}. \tag{S3}$$

Table S1 tabulates the Raman active modes according to the involved Wyckoff position. In Fig. 1a-c, show the atomic displacements for the three expected Ag modes obtained with the ISODISTORT [7, 8]. For the symmetric modes $A_{1g}$, neither the Fe atoms nor the Y atoms participate, as Table S1 shows. Fig. 1d and 1e show two examples of atomic displacements of the modes $E_g$, which involve Fe and Y atoms.

| | $A_{1g}$ | $E_g$ | $T_{2g}$ |
|---|---|---|---|
| Fe (16a) | x | x | x |
| Y (24c) | x | 1 | 2 |
| Fe (24d) | x | 1 | 3 |
| O (96h) | 3 | 6 | 9 |

Table S1. Wyckoff-position-resolved Raman active modes. The sum of each column corresponds to the total number of modes expected.

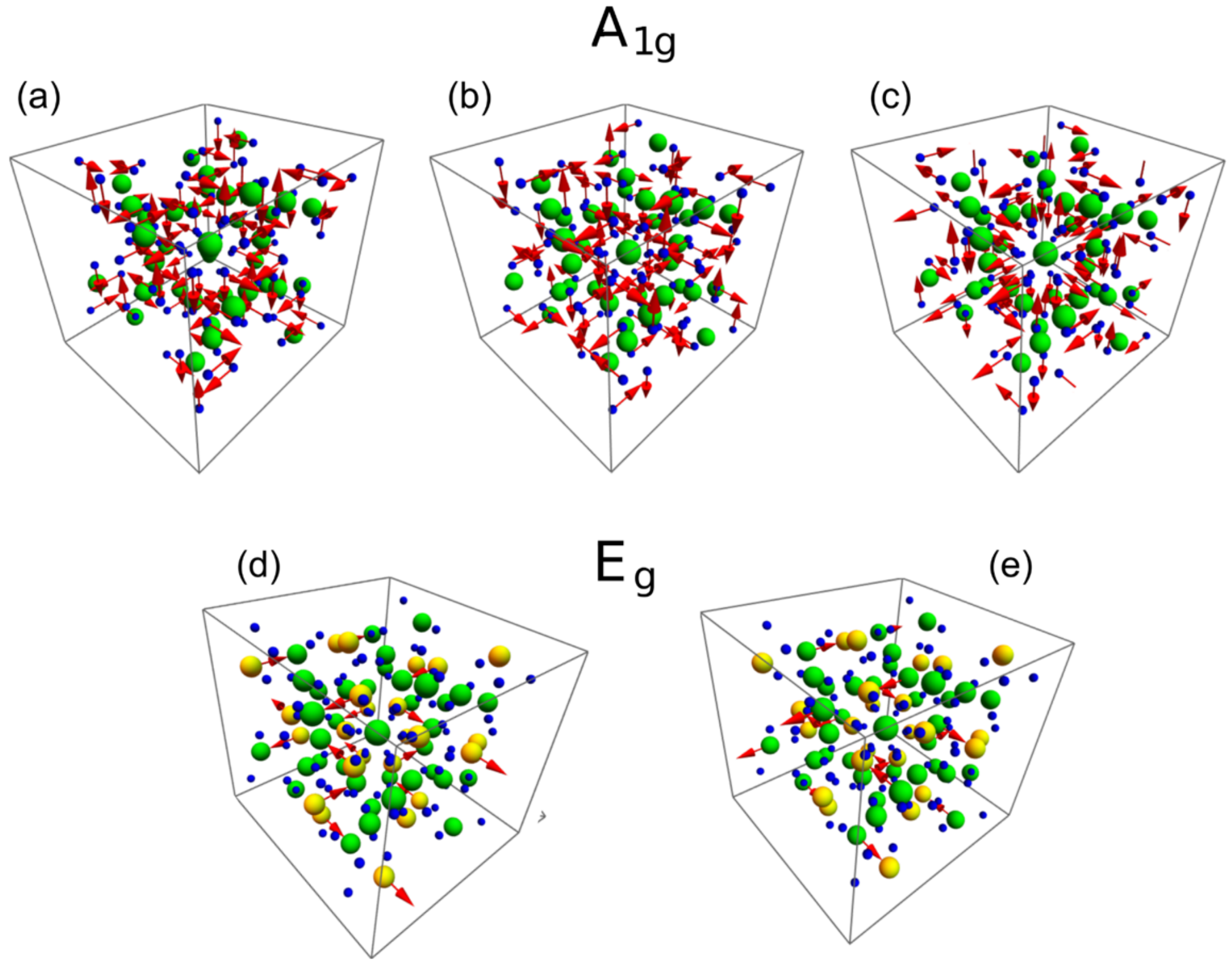

FIG. S1. (a)-(c) The three $A_{1g}$ Raman active modes in YIG viewed along the [111] direction. The O (96h) atoms are shown in blue, while the Fe (16a and 24d) are shown in green. For clarity, Y atoms (24c) are not shown. For the $A_{1g}$ modes, the Fe and Y atoms are stationary. (d) and (e) schematically show two of the eight $E_g$ irreducible representations in YIG along the [111] direction. Y atoms are shown in yellow. The mode displayed in (d) shows motion of the Y atoms, while (e) shows motion of the Fe atoms.

The Raman tensors, with respect to the principal axis of the crystal, are given by

$$R(A_{1g}) = \begin{pmatrix} a & 0 & 0 \\ 0 & a & 0 \\ 0 & 0 & a \end{pmatrix}, \quad R\left(E_g^{(1)}\right) = \begin{pmatrix} b & 0 & 0 \\ 0 & b & 0 \\ 0 & 0 & -2b \end{pmatrix}, \quad R\left(E_g^{(2)}\right) = \begin{pmatrix} -\sqrt{3}b & 0 & 0 \\ 0 & \sqrt{3}b & 0 \\ 0 & 0 & 0 \end{pmatrix},$$
$$R\left(T_{2g}^{(1)}\right) = \begin{pmatrix} 0 & 0 & 0 \\ 0 & 0 & d \\ 0 & d & 0 \end{pmatrix}, \quad R\left(T_{2g}^{(2)}\right) = \begin{pmatrix} 0 & 0 & d \\ 0 & 0 & 0 \\ d & 0 & 0 \end{pmatrix}, \quad R\left(T_{2g}^{(3)}\right) = \begin{pmatrix} 0 & d & 0 \\ d & 0 & 0 \\ 0 & 0 & 0 \end{pmatrix}. \tag{S4}$$

In order to carry out mode assignment in Raman measurements, it is useful to know the Raman scattering efficiency defined as

$$S \propto \left( \sum_{ij} e_{in}^{i} R_{ij} e_{s}^{j} \right)^2 \tag{S5}$$

where $e_{in/s}^{i}$ are the unit vectors for the incident and scattered polarization directions and $R$ is the Raman tensor [9]. For the degenerate modes $E_g$ and $T_{2g}$, we need to add the contributions from the two and three partner matrices, respectively.

The YIG sample used for the measurements has surface normal oriented along [111]. Therefore, we need to transform the Raman tensors such that the [111] direction corresponds to the z-direction in the new coordinate system. To determine the symmetry of the modes, we calculate the Raman intensity as a function of a rotation angle about the z-direction ([111]). In Fig. S2, we plot the Raman intensity for incident light parallel (perpendicular) to the scattered light.

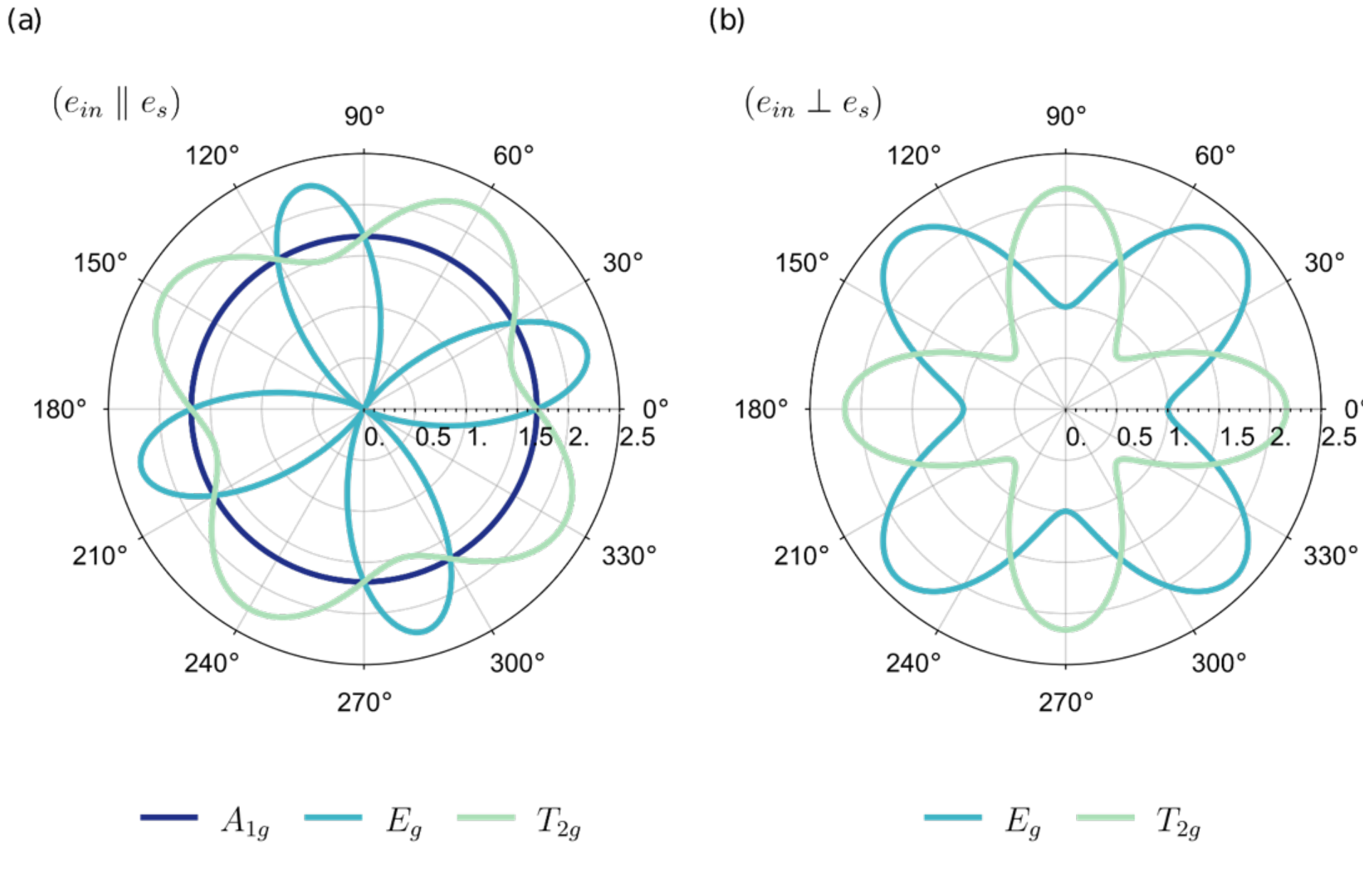

FIG. S2. Calculated Raman intensity as a function of sample rotation angle about the [111] direction for (a) parallel polarization and (b) perpendicular polarization.

## Section 2: Ginzburg-Landau Framework

This section describes in further detail the Ginzburg-Landau framework used to develop the mean-field model which appears in the main text.

### Section 2.1: Ferrimagnetic Order Ginzburg-Landau Potential

YIG's magnetic moment exhibits the temperature dependence shown in Fig. S3 as measured in Ref. [10, 11]. The solid line corresponds to a mean-field like fit

$$\frac{M(T)}{M_0} = \sqrt{1 - T/T_C}. \tag{S6}$$

The simple mean-field temperature dependences give a good first-order approximation. Therefore, we can describe the magnetic transition with the simple Ginzburg-Landau potential

$$F = \frac{A}{2}m^2 + \frac{B}{4}m^4, \tag{S7}$$

where $m \equiv M/M_0$ is the dimensionless ferrimagnetic order parameter oriented in the [111] direction, with $M_0 = 4.19361 \times 10^{-4}\mathrm{JT}^{-1}/\mathrm{cm}^3$. The GL parameters $A = -a(Tc - T)$and $B$ have units of energy.

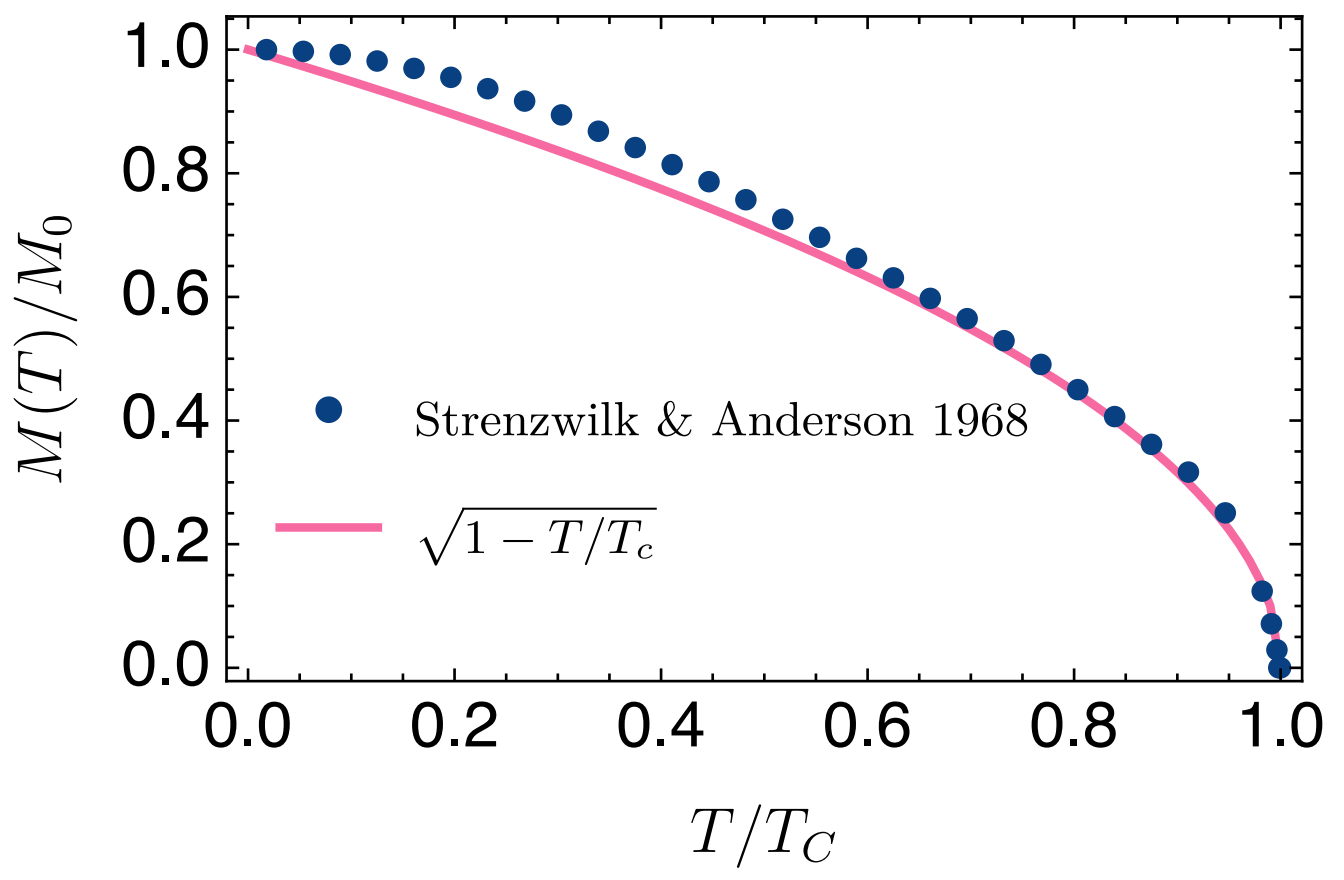

FIG. S3. YIG magnetic moment themperature dependence, as reported in Ref. [10], compared to the mean-field description with $T_C$ = 559 K.

### Section 2.2: Spin-Phonon Interaction

In this section ISOTROPY is used to obtain invariant polynomials for constructing the Ginzburg-Landau potential to describe the coupling between phonons and the magnetic order parameters [7]. Since the magnetic order parameter $m$ breaks time-reversal symmetry, only even powers are allowed in the GL potential. For the phonon modes, we consider only harmonic contributions. Similar approaches have been employed before to describe spin-phonon coupling in $YMnO_3$ [12] and $MnV_2O_4$ [13] in combination with first-principles calculations.

For each of the phonon symmetries, the Ginzburg-Landau potential takes the form

$$F_{A_{1g}} = F_0 + \frac{A}{2}m^2 + \frac{B}{2}m^4 + \frac{1}{2}\mu_A\omega_A u^2 + \frac{1}{2}m^2(\delta_A^L u + \delta_A^S u^2) \tag{S8}$$

$$F_{E_g} = F_0 + \frac{A}{2}m^2 + \frac{B}{2}m^4 + \frac{1}{2}\mu_E\omega_E u^2 + \frac{1}{2}\delta_E^S m^2 u^2 \tag{S9}$$

$$F_{T_{2g}} = F_0 + \frac{A}{2}m^2 + \frac{B}{2}m^4 + \frac{1}{2}\mu_T\omega_T u^2 + \frac{1}{2}m^2(\delta_T^L u + \delta_T^S u^2) \tag{S10}$$

where $\omega_i$ and $u_i$ are the bare frequency and reduced mass of the phonon modes. As determined in the first section, $A_{1g}$ modes only involve displacement of oxygen atoms, so $\mu_A = \mu_O$ is the oxygen mass. We use first-principles calculations of the Raman phonons to assign $\mu_E$ or $\mu_T$ to the phonon modes with symmetries $E_g$ and $T_{2g}$ in the next section. Finally, $\delta_i^L$ and $\delta_i^S$ correspond to the temperature-independent spin-phonon coupling constants, reported here in units of energy per Å and Å$^2$, respectively. In oxides, the coupling strength is usually of the order of a few $cm^{-1}$ [14]. The phonon modes with symmetries $E_g$ and $T_{2g}$ exhibit different coupling to the magnetic order.

The phonon modes with symmetries $E_g$ and $T_{2g}$ exhibit different coupling to the magnetic order. We now explore the implications of this in the temperature-dependence of the phonon frequency. The equilibrium values for the order parameters $u_*$ and $m_*$ are determined by the conditions

$$\frac{\partial F}{\partial u} = 0 \text{ and } \frac{\partial F}{\partial m} = 0, \tag{S11}$$

while the phonons frequency ($\Omega$) in the presence of coupling with spin is given by [12]

$$\mu\Omega^2 = \left.\frac{\partial^2 F}{\partial u^2}\right|_{\substack{u=u_* \\ m=m_*}} = \mu\omega^2 + \delta^S m_*^2 \tag{S12}$$

in the harmonic approximation for the three symmetry modes considered. Therefore, phonon modes with smaller reduced mass are expected to be more sensitive to magnetic-ordering effects than heavier phonon modes.

The frequency shift begins by evaluating the equilibrium conditions, yielding $m_* = \sqrt{a(T_c - T)/B}$. For the $E_g$ modes first, the temperature-dependent phonon frequency is

$$\Omega_E(T) = \sqrt{\omega_E^2 + \frac{\delta_E^S}{\mu}\frac{a}{B}(T_c - T)} \approx \omega + \frac{\delta_E^S}{2\mu\omega}\left(1 - \frac{T}{T_c}\right). \tag{S13}$$

The only undefined parameter is the spin-phonon coupling constant $\delta_E^S$ , which has units of energy per Å$^2$. The reduced mass $\mu$ is measured in kg and the angular frequency $\omega$ in $cm^{-1}$. Near the critical point, this result agrees with the correction derived in Ref. [15] from an expansion of the spin Hamiltonian as a Taylor series around small phonon displacements.

For the case of $A_{1g}$ and $T_{2g}$ modes, in contrast to the $E_g$ modes, the cubic term $m^2\delta^L$is allowed. This term indicates the presence of a distortion (non-zero solution for the equilibrium position $u_0$), however, from experiments, we know that distortions in YIG are very weak [16]. Therefore, we consider the limit $\delta^L \ll \delta^S$. Then, the same relation is obtained for the $A_{1g}$ and $T_{2g}$ modes,

$$\Omega(T) \approx \omega + \frac{\delta^S}{2\mu\omega}\left(1 - \frac{T}{T_c}\right). \tag{S14}$$

Eq. S14 is the mean-field equation used to fit the anharmonic subtracted phonons frequencies ($\Delta\omega$) to extract the spin-phonon interaction strength $\delta_{sp}$ in the main text.

## Section 3: First-principles evaluation of the Raman phonons

Density functional theory (DFT) calculations were used to calculate the Raman phonon frequencies and masses. We have used the Vienna ab intio Simulation Package (VASP) with projector-augmented waves pseudopotentials [17, 18]. The generalized gradient approximation is used as implemented in the Perdew-Burke-Ernzerhof functional revised for solids with an effective Hubbard term U-J=3.7 eV implemented in the Dudarev method [19, 20]. An energy convergence threshold of $10^{-6}$ eV was used. Structural parameters were relaxed using $10^{-3}$ eV/ Å force convergence condition on each atom. A $6 \times 6 \times 6$ k-point grid and 550 eV energy cut-off were found to give converged structural parameters.

Phonons were calculated at the $\Gamma$–point with the PhonoPy software [21]. Phonon masses were found by diagonalizing the generalized eigenvalue problem $\mu u = Ku$ where $K$ is the force-constant matrix and $\mu$ is the mass matrix.

The converged lattice constant is $a = 12.31$ Å with the free paramters of the O Wyckoff site converged to $x = 0.05763$, $y = 0.34949$, and $z = 0.47321$. The magnetic moments on the inequivalent $a$ and $d$ Fe sites were found to be 4.013 Bohr and -3.875 Bohr, respectively. The calculated Raman phonon frequencies are shown in Table S2 and change only quantitatively with reasonable variations in $U - J$. In Table S2 the measured phonon modes are assigned to a calculated $\mu$, based on the measured mode's symmetry and frequency. Note that the measured mode at 508 $cm^{-1}$ with symmetry $E_g$, $\omega_{508}$, does not have a well matched calculated mode. Previous measurements of YIG have located two modes close to this frequency, one with a $A_{1g}$ and another with $E_g$ symmetry [22]. Based on this, $\omega_{508}$ is matched to an $A_{1g}$ mode close to it's frequency instead of an $E_g$ mode. All the calculated modes within in 50 $cm^{-1}$ of $\omega_{508}$ are very similar in mass, thus this assignment does not drastically change the measured SPI.

| Symmetry | Phonon Frequency ($cm^{-1}$) | | $\mu$ | Symmetry | Phonon Frequency ($cm^{-1}$) | | $\mu$ |
|---|---|---|---|---|---|---|---|
| | DFT Calculated | Measured | | | DFT Calculated | Measured | |
| $A_{1g}$ | 333 | | 16.00 | $T_{2g}$ | 128 | | 62.29 |
| | 493 | 508* | 16.00 | | 168 | | 31.19 |
| | 713 | 739 | 16.00 | | 175 | 174 | 37.50 |
| $E_g$ | 130 | | 48.13 | | 189 | | 55.56 |
| | 267 | 276 | 19.87 | | 229 | | 21.42 |
| | 287 | | 24.02 | | 270 | 276 | 24.54 |
| | 336 | 346 | 16.54 | | 315 | | 18.37 |
| | 400 | | 18.51 | | 371 | | 16.71 |
| | 446 | | 16.19 | | 382 | 378 | 16.83 |
| | 624 | | 18.93 | | 429 | 447 | 16.18 |
| | 659 | | 16.04 | | 489 | | 17.10 |
| | | | | | 586 | | 18.11 |
| | | | | | 591 | 591 | 18.71 |
| | | | | | 719 | | 17.87 |

Table S2. Phonon symmetry, frequency, and mass for Raman phonons from DFT, with measured frequencies corresponding to the masses used in the calculating $\delta$. *The measured mode $\omega_{508}$ is matched to an $A_{1g}$ instead of an $E_g$, as no calculated $E_g$ modes are close to this frequency.

## Section 4: Anharmonic and Mean-Field Fits for All Phonon Modes

This section provides the anharmonic fits (Fig. S4) and mean-field fits (Fig. S5) for all of the measured phonon modes reported in the main text Table 1.

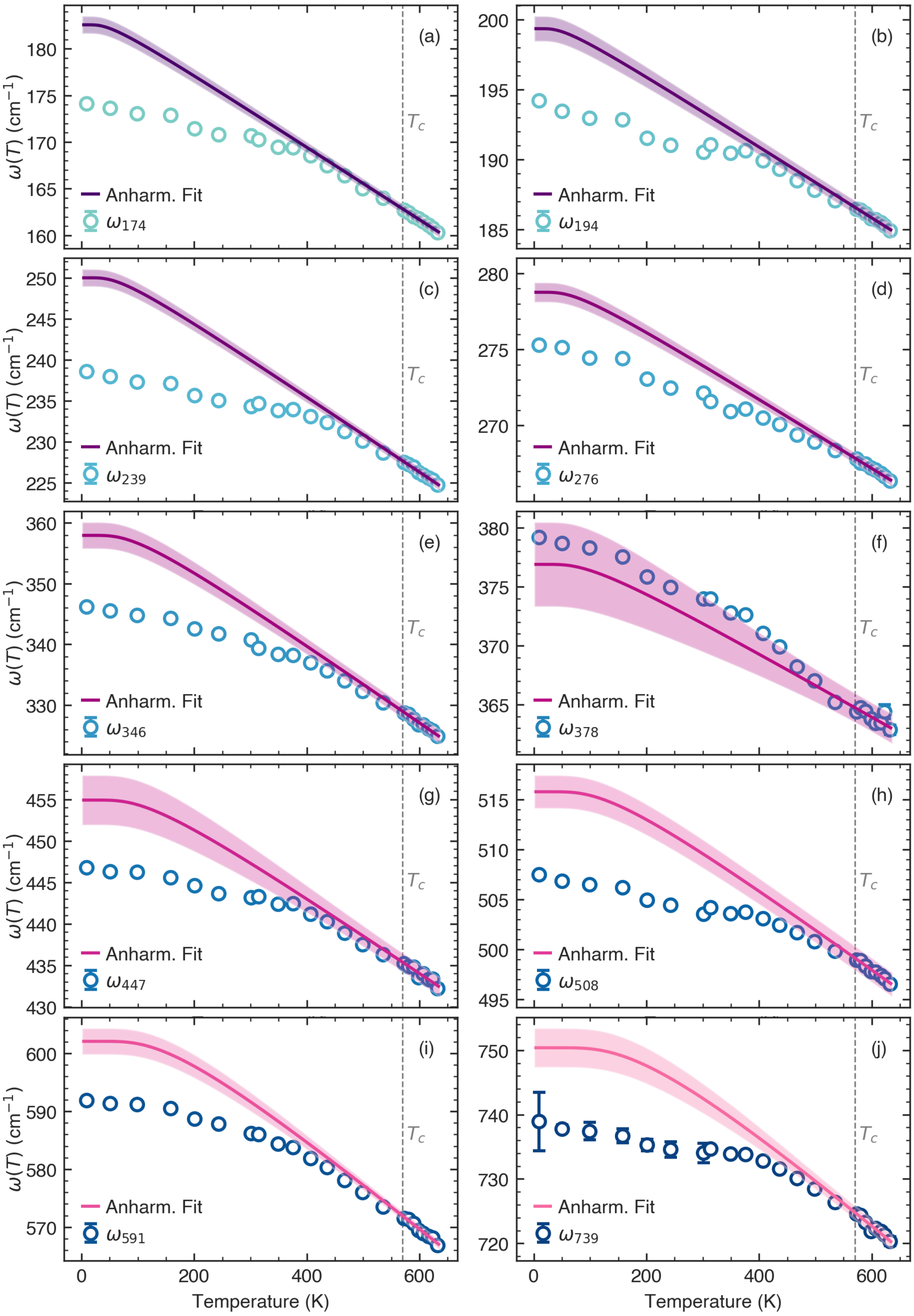

FIG. S4. Temperature dependent phonon frequencies extracted from Lorentzian fits for the 10 measured phonon modes. The anharmonic fit is performed using main text Eq. 1 on the frequencies above $T_c$. The difference between the fit and lowest temperature data point is used to determine $\Delta\omega_{sp}^0$ reported in the main text Table 1.

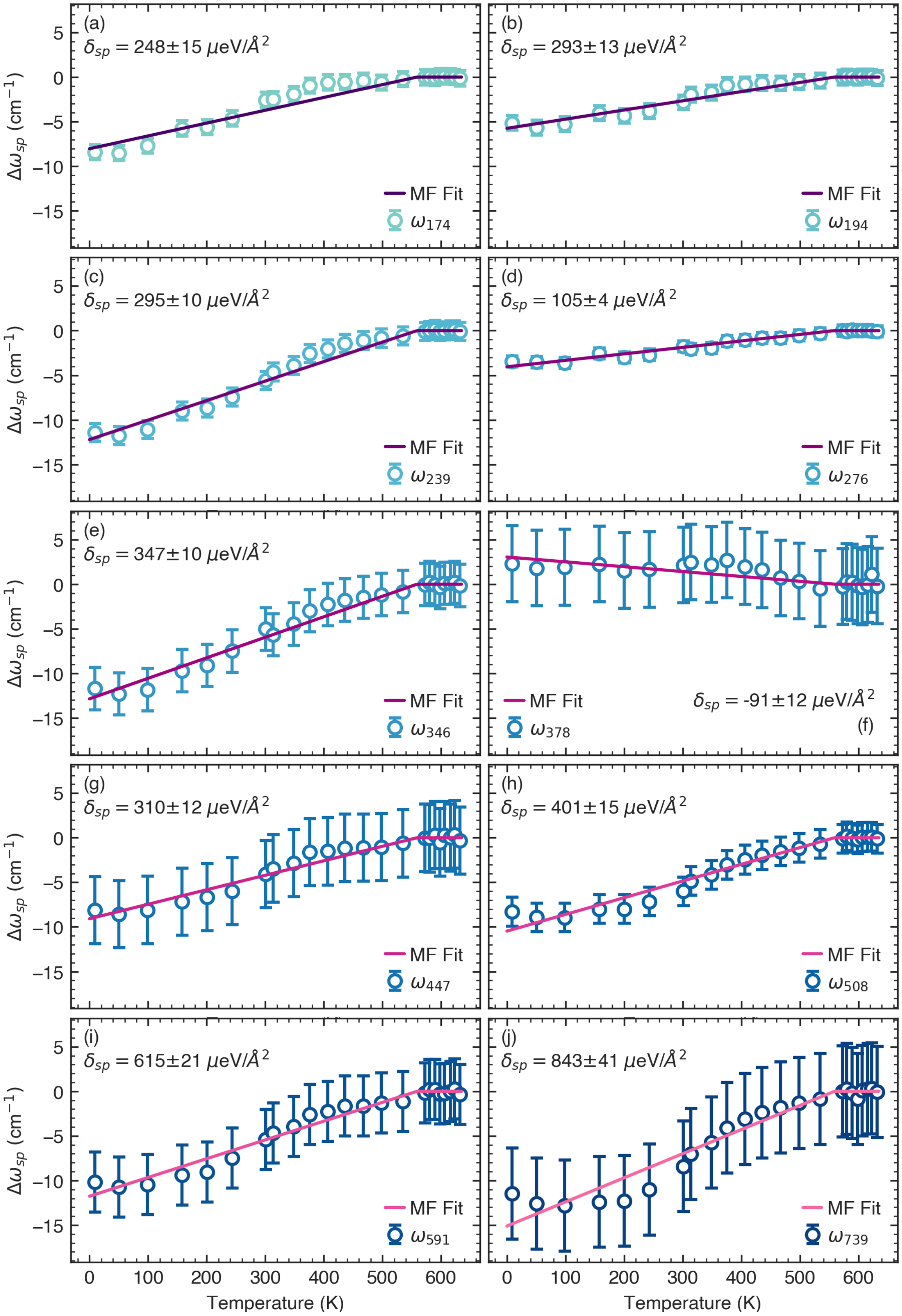

FIG. S5. Temperature dependent phonon frequency deviations found from subtracting data from the anharmonic fits shown in Fig. S3. The spin-phonon interaction strength $\delta_{sp}$ is found from fitting the mean-field model Eq. S14 to the data points. These are the $\delta_{sp}$ shown in Fig. 4 of the main text.

**Supplementary References:**

[1] S. Geller and M. A. Gilleo, "The crystal structure and ferrimagnetism of yttrium-iron garnet, Y3Fe2(FeO4)3," *J. Phys. Chem. Solids*, vol. **3**, no. 1–2, pp. 30–36, Jan. 1957.

[2] S. Geller and M. A. Gilleo, "The effect of dispersion corrections on the refinement of the yttrium-iron garnet structure," *J. Phys. Chem. Solids*, vol. **9**, no. 3–4, pp. 235–237, Mar. 1959.

[3] G. P. Rodrigue, H. Meyer and R. V. Jones, "Resonance measurements in magnetic garnets," *J. Appl. Phys.*, vol. **31**, no. 5, pp. S376–S382, May 1960.

[4] A. B. Harris, "Spin-wave spectra of yttrium and gadolinium iron garnet," *Phys. Rev.*, vol. **132**, no. 6, pp. 2398–2409, Dec. 1963.

[5] M. Wu, "Nonlinear spin waves in magnetic film feedback rings," in *Solid State Physics - Advances in Research and Applications*, vol. **62**, Academic Press, 2010, pp. 163–224.

[6] E. Kroumova, M. L. Aroyo, J. M. Perez-Mato, A. Kirov, C. Capillas, S. Ivantchev and H. Wondratschek, "Bilbao Crystallographic Server: Useful databases and tools for phase-transition studies," *Phase Transitions*, vol. **76**, no. 1–2, pp. 155–170, Jan. 2003.

[7] H. T. Stokes, D. M. Hatch and B. J. Campbell, "ISOTROPY Software Suite, iso.byu.edu." .

[8] B. J. Campbell, H. T. Stokes, D. E. Tanner and D. M. Hatch, "ISODISPLACE: A web-based tool for exploring structural distortions," *J. Appl. Crystallogr.*, vol. **39**, no. 4, pp. 607–614, Aug. 2006.

[9] R. Loudon, "The Raman effect in crystals," *Adv. Phys.*, vol. **13**, no. 52, pp. 423–482, 1964.

[10] D. F. Strenzwilk and E. E. Anderson, "Calculation of the sublattice magnetization of yttrium iron garnet by the Oguchi method," *Phys. Rev.*, vol. **175**, no. 2, pp. 654–659, Nov. 1968.

[11] E. E. Anderson, "Molecular field model and the magnetization of YIG," *Phys. Rev.*, vol. **134**, no. 6A, p. A1581, Jun. 1964.

[12] A. Paul, P. Sharma and U. V. Waghmare, "Spin-orbit interaction, spin-phonon coupling, and anisotropy in the giant magnetoelastic effect in YMnO3," *Phys. Rev. B*, vol. **92**, no. 5, p. 054106, Aug. 2015.

[13] D. Dey, T. Maitra, U. V. Waghmare and A. Taraphder, "Phonon dispersion, Raman spectra, and evidence for spin-phonon coupling in MnV2 O4 from first principles," *Phys. Rev. B*, vol. **101**, no. 20, p. 205132, May 2020.

[14] A. B. Sushkov, O. Tchernyshyov, W. Ratcliff, S. W. Cheong and H. D. Drew, "Probing spin correlations with phonons in the strongly frustrated magnet ZnCr2O4," *Phys. Rev. Lett.*, vol. **94**, no. 13, p. 137202, Apr. 2005.

[15] E. Granado, A. García, J. A. Sanjurjo, C. Rettori, I. Torriani, F. Prado, R. D. Sánchez, A. Caneiro and S. B. Oseroff, "Magnetic ordering effects in the Raman spectra of La1-xMn1-xO3," *Phys. Rev. B*, vol. **60**, no. 17, pp. 11879–11882, Nov. 1999.

[16] R. Z. Levitin, A. S. Markosyan and V. N. Orlov, "X-ray study of crystalline structure magnetoelastic distortions in Tb3−xYxFe5O12 terbium-yttrium ferrite-garne," *Sov. Phys. Solid State*, vol. **25**, no. 6, pp. 1074–1075, 1983.

[17] G. Kresse and J. Furthmüller, "Efficient iterative schemes for ab initio total-energy calculations using a plane-wave basis set," *Phys. Rev. B*, vol. **54**, no. 16, pp. 11169–11186, Oct. 1996.

[18] G. Kresse and D. Joubert, "From ultrasoft pseudopotentials to the projector augmented-wave method," *Phys. Rev. B*, vol. **59**, no. 3, pp. 1758–1775, Jan. 1999.

[19] S. Dudarev and G. Botton, "Electron-energy-loss spectra and the structural stability of nickel oxide: An LSDA+U study," *Phys. Rev. B*, vol. **57**, no. 3, pp. 1505–1509, Jan. 1998.

[20] J. P. Perdew, K. Burke and M. Ernzerhof, "Generalized gradient approximation made simple," *Phys. Rev. Lett.*, vol. **77**, no. 18, pp. 3865–3868, Oct. 1996.

[21] A. Togo and I. Tanaka, "First principles phonon calculations in materials science," *Scr. Mater.*, vol. **108**, pp. 1–5, Nov. 2015.

[22] W. H. Hsu, K. Shen, Y. Fujii, A. Koreeda and T. Satoh, "Observation of terahertz magnon of Kaplan-Kittel exchange resonance in yttrium-iron garnet by Raman spectroscopy," *Phys. Rev. B*, vol. **102**, no. 17, p. 174432, Nov. 2020.